\pdfoutput=1
\documentclass[12pt,a4paper]{article}

\usepackage[margin=1in]{geometry}
\usepackage{amsmath,amssymb}
\usepackage{graphicx}
\usepackage{booktabs}
\usepackage{array}
\usepackage{float}
\usepackage{caption}
\usepackage{hyperref}
\usepackage{xcolor}
\usepackage{microtype}
\usepackage{setspace}
\usepackage[numbers,sort&compress]{natbib}

\graphicspath{{figures/}}
\hypersetup{
    colorlinks=true,
    linkcolor=blue,
    citecolor=blue,
    urlcolor=blue
}

\title{Phonon-Aware Electron-to-Nuclear Spin Quantum Memory in NV Centers}

\author{
Manjith B C\\
Department of Computer Science and Engineering -- Cyber Security\\
Indian Institute of Information Technology Kottayam\\
Kerala, India\\
\texttt{manjithbc@iiitkottayam.ac.in}
}

\date{}

\begin{document}

\maketitle

\begin{abstract}
Nitrogen-vacancy (NV) centers in diamond are a promising solid-state platform for quantum memory: the electron spin can be optically initialized and read out, while nearby nuclear spins have long coherence times and can serve as storage qubits. In this work, we develop a phonon-aware open quantum system model for electron-to-nuclear spin state transfer in a single NV center. The model includes the NV electron spin, a coupled \(^{13}\mathrm{C}\) nuclear spin, the hyperfine interaction, an effective microwave/radio-frequency-assisted SWAP interaction for writing and reading, a single effective phonon mode with longitudinal spin-phonon coupling, and Lindblad relaxation and dephasing channels. The dynamics are simulated with the Lindblad master equation in QuTiP, using the secular approximation and a hyperfine-resonant drive in the rotating frame. Without spin-phonon coupling and electron decoherence, the write--store--read protocol recovers the stored state with a fidelity above 0.9999. With the default decoherence rates, increasing the spin-phonon coupling from 0 to 0.06\,MHz lowers the worst-case retrieval fidelity from 0.9994 to 0.9884, and the fidelity oscillates with the phonon-mode period because electron--phonon entanglement generated during the write step is partially undone during the read step. With a phenomenological temperature-dependent rate model, the fidelity decreases monotonically from 0.9997 at 1\,K to 0.9938 at 300\,K, while storage-time losses follow the assumed nuclear-spin decoherence rates. The storage-time plots, temperature sweep, and parameter maps identify operating regimes for NV-center quantum memory, and the accompanying open-source code provides a reproducible framework for studying phonon-limited quantum memory.
\end{abstract}


\section{Introduction}

\begin{figure}[htbp]
  \centering
  \includegraphics[width=0.85\textwidth]{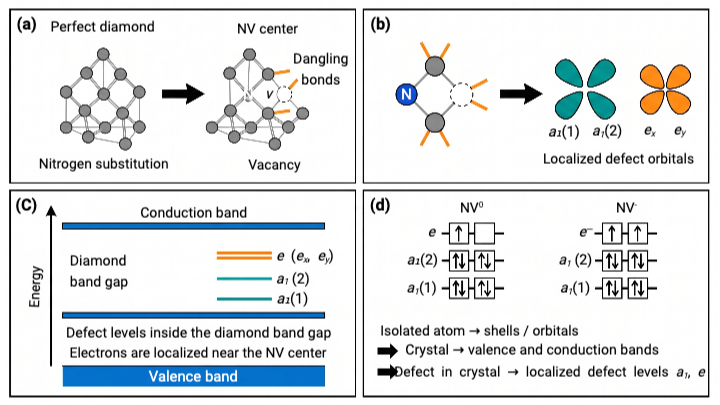}
  \caption{Conceptual overview of the NV center: (a) formation from a substitutional nitrogen atom and an adjacent vacancy, (b) localized defect orbitals, (c) defect levels within the diamond band gap, and (d) orbital occupations. This illustration was generated with an AI image-generation tool and is included for conceptual orientation only; it is not used for any quantitative result.}
\end{figure}

Nitrogen-vacancy (NV) centers in diamond are widely studied solid-state quantum systems for applications in quantum information processing, quantum sensing, and quantum networks. An NV center is a defect in the diamond lattice formed by a substitutional nitrogen atom next to a vacant carbon site. The negatively charged NV center, denoted as \(\mathrm{NV}^{-}\), is especially important because it possesses an electronic spin triplet ground state that can be used as a controllable quantum system. The electron spin of the \(\mathrm{NV}^{-}\) center can be initialized optically, manipulated coherently using microwave fields, and read out through spin-dependent fluorescence. These properties make the NV electron spin a practical qubit for quantum operations. In addition, NV centers can maintain useful spin coherence even under relatively accessible experimental conditions, including room temperature, which makes them attractive compared with many other solid-state quantum platforms \cite{doherty2013,jelezko2006,childress2013}.

Although the NV electron spin is highly useful for initialization, control, and readout, it is also more directly exposed to environmental noise. Interactions with magnetic-field fluctuations, surrounding spin impurities, charge noise, and lattice vibrations can reduce the coherence time of the electron spin. For long-term storage of quantum information, nearby nuclear spins offer an important advantage. Nuclear spins couple more weakly to the environment and therefore generally exhibit longer coherence times than electron spins. In NV-based quantum memory and quantum network architectures, the electron spin can serve as an active communication or processing qubit, while nearby nuclear spins, such as intrinsic nitrogen nuclear spins or surrounding \(^{13}\mathrm{C}\) nuclear spins, can serve as long-lived memory qubits. In such a system, a quantum state can be prepared and controlled using the electron spin and then transferred to a nuclear spin for storage \cite{fuchs2011,shim2013,bradley2019}.

Many basic models of NV-center quantum memory focus mainly on the hyperfine interaction between the electron spin and the nuclear spin. These models often include standard relaxation and dephasing terms to represent decoherence. However, such simplified approaches may not fully capture the role of phonons in limiting memory performance. Phonons are quantized lattice vibrations in the diamond crystal, and they can influence the NV spin through spin-strain coupling and spin-lattice relaxation processes. Mechanical vibrations and lattice strain can modify the local environment of the NV center and disturb the electron spin states. Temperature-dependent spin-lattice relaxation also shows that phonon-related processes can reduce spin coherence, especially during electron-spin-based control and readout operations. Therefore, a phonon-aware model is important for understanding how lattice vibrations affect electron-to-nuclear spin state transfer and quantum memory fidelity \cite{macquarrie2013,cambria2023,thiering2025}.

This paper develops a phonon-aware electron-to-nuclear spin quantum memory model for NV centers in diamond. The proposed model includes the NV electron spin, a nearby nuclear-spin memory qubit, hyperfine-mediated state transfer, and phonon-induced relaxation and dephasing effects. The main objective is to evaluate how phonon-induced electron-spin decoherence affects the transfer of a quantum state from the electron spin to a nuclear spin and its later retrieval. By analyzing memory fidelity under different phonon coupling strengths, relaxation rates, and storage times, this work provides a simulation-based framework for studying phonon effects in NV-center quantum memories.

The main contributions of this work are summarized as follows:
\begin{enumerate}
\item \textbf{Development of a phonon-aware quantum memory model:} We develop an open quantum system model for electron-to-nuclear spin quantum memory in NV centers that explicitly incorporates phonon-induced effects during quantum state transfer and storage.
\item \textbf{Unified Hamiltonian framework:} We formulate a comprehensive Hamiltonian that combines the NV electron spin, nuclear spin, electron--nuclear hyperfine interaction, phonon modes, spin-phonon coupling, and external control fields within a single simulation framework.
\item \textbf{Open-system simulation using the Lindblad master equation:} We model realistic quantum memory operation by incorporating electron-spin relaxation, electron-spin dephasing, nuclear-spin decoherence, and phonon damping through the Lindblad master equation \cite{lindblad1976,gorini1976,breuer2002}.
\item \textbf{Quantitative analysis of phonon-induced decoherence:} We investigate how phonon-induced relaxation and dephasing influence the fidelity of electron-to-nuclear quantum state transfer, storage, and retrieval under different operating conditions.
\item \textbf{Performance evaluation under varying physical parameters:} We evaluate the impact of spin-phonon coupling strength, storage time, relaxation rates, and temperature-dependent phonon occupation on quantum memory performance.
\item \textbf{Design insights for robust NV-center quantum memory:} Based on the simulation results, we identify operating conditions that minimize phonon-induced decoherence and improve memory fidelity, providing useful guidelines for the design of future NV-center-based quantum memories.
\item \textbf{A reproducible simulation framework:} The proposed QuTiP-based framework \cite{johansson2012,johansson2013} can be readily extended to investigate additional decoherence mechanisms, multiple nuclear-spin registers, quantum error-mitigation techniques, and hybrid quantum memory architectures.
\end{enumerate}

\section{Related Work}

\subsection{NV-Center Electron and Nuclear Spin Systems}

The spin system of a negatively charged nitrogen-vacancy center is mainly described by its electronic spin triplet ground state with spin quantum number \(S=1\) \cite{doherty2013,jelezko2006}. In the absence of an external magnetic field, the ground-state sublevels are separated by the zero-field splitting parameter \(D\), which separates the \(m_s=0\) state from the \(m_s=\pm 1\) states. This energy separation allows two of the spin sublevels, commonly \(m_s=0\) and \(m_s=-1\), to be selected as an effective two-level qubit system. The electronic spin Hamiltonian is commonly written in terms of spin operators \(S_x\), \(S_y\), and \(S_z\), together with zero-field splitting and Zeeman terms. This Hamiltonian description is important because it provides the basis for understanding microwave-driven spin transitions, energy-level tuning using magnetic fields, and coherent control of the NV electron spin.

The NV center is also naturally coupled to nuclear spins located near the defect. These include the intrinsic nitrogen nuclear spin and nearby \(^{13}\mathrm{C}\) nuclear spins in the diamond lattice. The intrinsic nitrogen nucleus may be \(^{14}\mathrm{N}\), with nuclear spin \(I=1\), or \(^{15}\mathrm{N}\), with nuclear spin \(I=1/2\), depending on the isotope. In addition, naturally occurring \(^{13}\mathrm{C}\) nuclei have nuclear spin \(I=1/2\), while the more abundant \(^{12}\mathrm{C}\) nuclei have no nuclear spin. Because of this, \(^{13}\mathrm{C}\) nuclei near the NV center can act as individually addressable nuclear-spin qubits. The number, position, and coupling strength of nearby \(^{13}\mathrm{C}\) nuclei strongly influence the structure of the NV-based quantum register \cite{childress2013,bradley2019}.

The interaction between the NV electron spin and a nearby nuclear spin is mainly described by the hyperfine interaction \cite{childress2013,fuchs2011,shim2013}. This interaction depends on the relative position of the nuclear spin with respect to the NV center and can contain both parallel and transverse coupling components. The parallel hyperfine term shifts the nuclear-spin transition frequency depending on the electron-spin state, while the transverse hyperfine term can enable electron--nuclear spin mixing and state-transfer processes under suitable control conditions. Therefore, the hyperfine interaction is the physical mechanism that links the controllable electron spin to the nuclear-spin memory qubit.

In NV-center quantum registers, the electron spin is usually treated as the central controllable qubit, while nearby nuclear spins are treated as auxiliary memory qubits. The electron spin provides fast control, optical initialization, and optical readout, whereas the nuclear spins provide long-lived storage. This combination allows the construction of a hybrid spin register in which quantum information can be transferred between electron and nuclear degrees of freedom. Strongly coupled nuclear spins are useful for fast gate operations, while weakly coupled nuclear spins can provide longer storage times but require more careful control. Thus, the electron--nuclear spin system of the NV center provides a flexible platform for realizing small-scale quantum registers and quantum memories.

From a modeling perspective, the complete NV electron--nuclear spin system can be represented using a composite Hilbert space formed by the tensor product of the electron-spin and nuclear-spin subspaces. The total Hamiltonian includes the electron-spin Hamiltonian, nuclear Zeeman Hamiltonian, hyperfine interaction, and externally applied microwave or radio-frequency control fields. This framework makes it possible to simulate coherent electron-spin control, nuclear-spin manipulation, and electron-to-nuclear quantum state transfer. Such Hamiltonian-based modeling is the foundation for analyzing NV-center quantum memory performance under both ideal and noisy conditions.

\subsection{Electron-to-Nuclear Spin Quantum Memory}

Electron-to-nuclear spin state transfer is a key mechanism for realizing quantum memory in NV centers. The basic idea is to first prepare or receive a quantum state in the NV electron spin and then transfer this state to a nearby nuclear spin for storage. The electron spin is useful because it can be initialized and measured efficiently, but it is also more sensitive to environmental noise. The nuclear spin, on the other hand, is more weakly coupled to the environment and can preserve quantum information for a longer time. Therefore, the electron spin can act as an interface qubit, while the nuclear spin acts as a memory qubit.

A significant early demonstration was reported by Fuchs et al. \cite{fuchs2011}, where the electronic spin state of a single NV center was coherently stored in the intrinsic nitrogen nuclear spin at room temperature. Their work used hyperfine-mediated avoided level crossings and Landau--Zener transitions to transfer quantum information between the electron spin and the nitrogen nuclear spin. This demonstrated that the nuclear spin naturally present in each NV center can function as an intrinsic quantum memory.

In addition to intrinsic nitrogen nuclear spins, \(^{13}\mathrm{C}\) nuclear spins in the surrounding diamond lattice have also been explored as memory qubits. Shim et al. demonstrated a room-temperature high-speed nuclear-spin quantum memory using a single \(^{13}\mathrm{C}\) nuclear spin strongly coupled to the NV electron spin \cite{shim2013}. The strong hyperfine interaction enabled fast transfer between the electron-spin register qubit and the \(^{13}\mathrm{C}\) nuclear-spin memory qubit, showing the usefulness of nearby carbon nuclear spins for fast quantum memory operations.

For quantum-network applications, nuclear-spin memories are especially important because the electron spin is often involved in optical communication or entanglement generation. Reiserer et al. investigated an NV-based network node where the electron spin was hyperfine-coupled to nearby nuclear spins and showed that nuclear-spin coherence can be protected during repeated optical network operations \cite{reiserer2016}. Bradley et al. later demonstrated robust NV-center spin registers and a \(^{13}\mathrm{C}\)-spin quantum memory in isotopically engineered diamond that could tolerate optical link operations of the NV center, showing the relevance of nuclear-spin memories for scalable quantum networks \cite{bradley2019,bradley2022}.

\subsection{Spin-Phonon and Spin-Lattice Effects}

Although many NV-center memory models focus mainly on electron--nuclear hyperfine interaction, relaxation, and dephasing, phonon-related effects are also important. Phonons are quantized vibrations of the diamond lattice, and they can influence the NV spin through strain, spin-phonon coupling, and spin-lattice relaxation. Since NV-center quantum memory relies on coherent spin control and state transfer, phonon-induced disturbances can reduce the fidelity of write, storage, and read operations.

Direct coupling between phonons and NV-center spins has been experimentally demonstrated using mechanically generated strain \cite{macquarrie2013}. MacQuarrie et al. showed that mechanical lattice vibrations can drive spin transitions in NV centers, providing evidence that phonons can directly interact with the NV spin system. This result is important because it shows that lattice vibrations are not merely a background effect, but can actively couple to the spin degrees of freedom.

Temperature-dependent spin-lattice relaxation further shows the importance of phonon effects in NV centers \cite{cambria2023}. Cambria et al. reported that spin-lattice relaxation within the NV electronic ground-state spin triplet limits coherence times and affects the performance of NV-based quantum applications. Their temperature-dependent measurements from 9 K to 474 K were explained using a theory involving second-order spin-phonon interactions and phonon-assisted relaxation processes. This indicates that phonon-induced relaxation can become an important limitation for high-fidelity NV quantum technologies.

Recent theoretical work has also extended the role of phonons to nuclear-spin relaxation in solid-state defect qubits \cite{thiering2025}. Thiering and Gali studied nuclear-spin relaxation through electron-phonon coupling in the optical excited state of the NV center. Their work suggests that electron-phonon interaction can indirectly influence nuclear-spin relaxation, which is particularly relevant for NV-center memory systems where nuclear spins are used as long-lived storage qubits.

Therefore, while existing NV quantum-memory studies have demonstrated electron-to-nuclear spin storage using intrinsic nitrogen and \(^{13}\mathrm{C}\) nuclear spins, the explicit effect of phonon-induced electron relaxation and dephasing on memory fidelity remains an important modeling direction. This motivates the present work, which develops a phonon-aware electron-to-nuclear spin quantum memory model for NV centers and evaluates how phonon-induced decoherence affects quantum-state transfer, storage, and retrieval.

\section{System Model and Hamiltonian}

\begin{figure}[htbp]
  \centering
  \includegraphics[width=\textwidth]{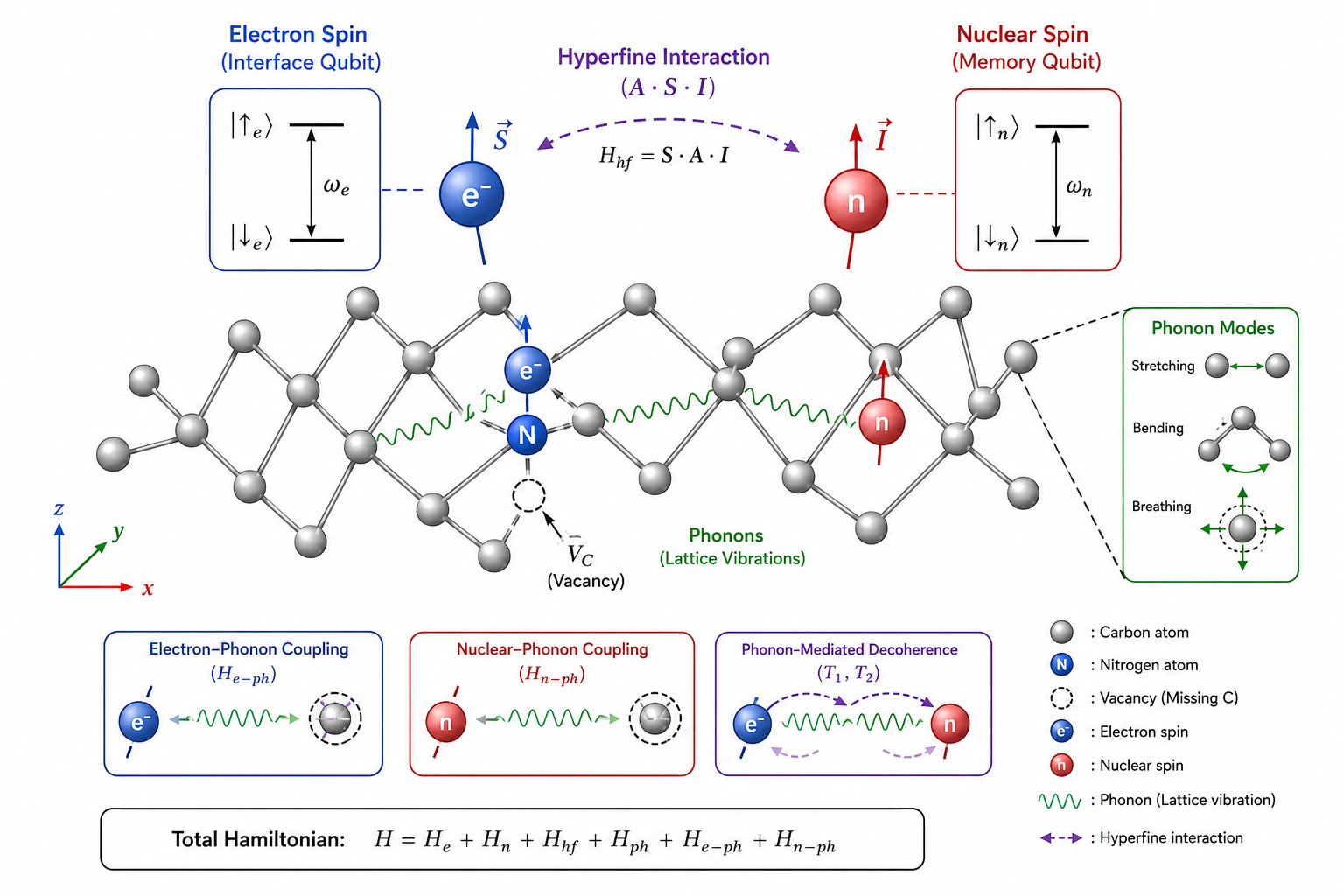}
  \caption{Phonon-aware electron-to-nuclear spin quantum memory architecture in an NV center. The diagram shows the NV electron spin as the interface qubit, the nearby nuclear spin as the memory qubit, the hyperfine interaction responsible for electron--nuclear state transfer, and phonon modes that introduce electron--phonon and nuclear--phonon coupling channels leading to phonon-mediated decoherence.}
  \label{fig:electron_nuclear_phonon_architecture}
\end{figure}

This section presents the theoretical model of the proposed phonon-aware electron-to-nuclear spin quantum memory in a nitrogen-vacancy center in diamond. The system consists of three main quantum components: the NV electron spin, a nearby nuclear spin used as the memory qubit, and an effective phonon mode representing lattice vibrations in the diamond crystal. The electron spin acts as the controllable interface qubit, while the nuclear spin acts as the long-lived storage qubit. The phonon mode is included to study how lattice vibrations influence the fidelity of electron-to-nuclear spin state transfer. Figure~\ref{fig:electron_nuclear_phonon_architecture} illustrates the overall architecture and the coupling pathways considered in the model.

The tensor-product structure follows the standard composite-system description used for coupled electron, nuclear, and bosonic modes in open quantum systems \cite{breuer2002,nielsen2010}. The total Hilbert space of the system is written as
\begin{equation}
\mathcal{H}=\mathcal{H}_{e}\otimes \mathcal{H}_{n}\otimes \mathcal{H}_{ph},
\end{equation}
where \(\mathcal{H}_{e}\) represents the NV electron-spin space, \(\mathcal{H}_{n}\) represents the nuclear-spin space, and \(\mathcal{H}_{ph}\) represents the phonon-mode space. For the negatively charged NV center, the electron spin has spin quantum number \(S=1\), with three ground-state spin sublevels \(m_s=0\) and \(m_s=\pm 1\). In the present memory model, the two levels \(\left|m_s=0\right\rangle\) and \(\left|m_s=-1\right\rangle\) are used as the effective electron-spin qubit. The nearby nuclear spin, such as a \(^{13}\mathrm{C}\) nuclear spin or an intrinsic nitrogen nuclear spin, is used as the quantum memory qubit.

Following standard NV-center spin-Hamiltonian descriptions and adding an effective phonon mode, the total Hamiltonian of the phonon-aware NV electron--nuclear spin system is expressed as
\begin{equation}
H=H_e+H_n+H_{hf}+H_{ph}+H_{e-ph}+H_{drive}.
\end{equation}
Here, \(H_e\) is the NV electron-spin Hamiltonian, \(H_n\) is the nuclear-spin Hamiltonian, \(H_{hf}\) is the electron--nuclear hyperfine interaction, \(H_{ph}\) is the phonon Hamiltonian, \(H_{e-ph}\) is the effective electron--phonon interaction, and \(H_{drive}\) represents the external microwave or radio-frequency control fields used for memory write and read operations \cite{doherty2013,childress2013,macquarrie2013}.

\subsection{NV Electron-Spin Hamiltonian}

The NV electron-spin Hamiltonian is written in the usual zero-field-splitting plus Zeeman form \cite{doherty2013,jelezko2006}:
\begin{equation}
H_e=D S_z^2+\gamma_e B_z S_z .
\end{equation}
In this expression, \(D\) is the zero-field splitting parameter of the NV center, \(S_z\) is the electron-spin operator along the NV symmetry axis, \(\gamma_e\) is the electron gyromagnetic ratio, and \(B_z\) is the applied magnetic field along the NV axis. The term \(D S_z^2\) separates the \(m_s=0\) state from the \(m_s=\pm 1\) states even in the absence of an external magnetic field. For NV centers, this zero-field splitting is approximately \(2.87\,\mathrm{GHz}\), which places the spin transition in the microwave frequency range.

The second term, \(\gamma_e B_z S_z\), represents the Zeeman interaction between the NV electron spin and the external magnetic field. This term lifts the degeneracy between the \(m_s=+1\) and \(m_s=-1\) states. By applying a suitable magnetic field, the transition between \(\left|m_s=0\right\rangle\) and \(\left|m_s=-1\right\rangle\) can be spectrally selected and used as an effective two-level electron-spin qubit. Therefore, the electron-spin qubit basis is defined as
\begin{equation}
\left|0\right\rangle_e \equiv \left|m_s=0\right\rangle,\qquad
\left|1\right\rangle_e \equiv \left|m_s=-1\right\rangle .
\end{equation}
This electron-spin qubit is suitable for initialization, microwave control, and optical readout, making it the active interface qubit in the proposed memory architecture \cite{doherty2013,jelezko2006}.

\subsection{Nuclear-Spin Hamiltonian}

The nuclear-spin Hamiltonian is represented by the nuclear Zeeman interaction \cite{childress2013}:
\begin{equation}
H_n=-\gamma_n B_z I_z .
\end{equation}
Here, \(\gamma_n\) is the nuclear gyromagnetic ratio and \(I_z\) is the nuclear-spin operator along the applied magnetic-field direction. The negative sign follows the usual convention for the nuclear Zeeman interaction. The nuclear spin may correspond to an intrinsic nitrogen nucleus or a nearby \(^{13}\mathrm{C}\) nucleus in the diamond lattice. For a \(^{13}\mathrm{C}\) nuclear spin, the spin quantum number is
\begin{equation}
I=\frac{1}{2},
\end{equation}
and the two nuclear-spin states may be represented as
\begin{equation}
\left|0\right\rangle_n,\qquad \left|1\right\rangle_n .
\end{equation}
These two states form the nuclear-spin memory qubit. Compared with the electron spin, the nuclear spin generally couples more weakly to environmental noise because its gyromagnetic ratio is much smaller. Therefore, the nuclear spin is suitable for storing quantum information over longer times. In the proposed memory model, the electron spin is used for fast control and readout, while the nuclear spin is used for long-lived storage.

\subsection{Electron--Nuclear Hyperfine Interaction}

The interaction between the NV electron spin and the nearby nuclear spin is described by the hyperfine Hamiltonian \cite{childress2013,fuchs2011,shim2013}:
\begin{equation}
H_{hf}=A_{\parallel}S_z I_z+A_{\perp}\left(S_x I_x+S_y I_y\right).
\end{equation}
Here, \(A_{\parallel}\) and \(A_{\perp}\) are the parallel and transverse hyperfine coupling strengths, respectively. The operators \(S_x\), \(S_y\), and \(S_z\) are the electron-spin operators, while \(I_x\), \(I_y\), and \(I_z\) are the nuclear-spin operators.

The first term,
\begin{equation}
A_{\parallel}S_z I_z,
\end{equation}
is the longitudinal or parallel hyperfine interaction. It produces a conditional energy shift depending on both the electron-spin projection and the nuclear-spin projection along the NV axis. For a combined electron--nuclear state \(\left|m_s\right\rangle_e\left|m_I\right\rangle_n\), this term contributes an energy proportional to
\begin{equation}
A_{\parallel}m_s m_I .
\end{equation}
Thus, the nuclear-spin transition frequency depends on the electron-spin state. This state-dependent shift enables conditional control of the nuclear spin and is important for constructing electron--nuclear quantum gates.

The second term,
\begin{equation}
A_{\perp}\left(S_x I_x+S_y I_y\right),
\end{equation}
is the transverse hyperfine interaction. It can be rewritten in terms of spin raising and lowering operators as
\begin{equation}
S_x I_x+S_y I_y=\frac{1}{2}\left(S_{+}I_{-}+S_{-}I_{+}\right).
\end{equation}
This form shows that the transverse hyperfine term can mediate spin exchange between the electron and nuclear spins. The term \(S_{-}I_{+}\) corresponds to a process in which the electron spin changes its state while the nuclear spin is raised, whereas \(S_{+}I_{-}\) corresponds to the reverse process. These exchange-like processes are central to electron-to-nuclear spin state transfer. In practice, because the electron and nuclear transition frequencies are very different, direct flip-flop transfer may require microwave or radio-frequency assistance, rotating-frame control, or operation near suitable resonance conditions.

Therefore, the hyperfine Hamiltonian provides the physical link between the controllable electron-spin qubit and the nuclear-spin memory qubit. Without this term, the electron and nuclear spins would evolve independently, and quantum-state transfer from the electron spin to the nuclear spin would not be possible.

\subsection{Phonon Hamiltonian}

To include lattice vibrations, the phonon mode is modeled as a quantized harmonic oscillator. This effective bosonic-mode description is commonly used for phonon and mechanical-mode modeling in open quantum systems \cite{breuer2002,macquarrie2013}. The phonon Hamiltonian is written as
\begin{equation}
H_{ph}=\hbar \omega_m a^{\dagger}a .
\end{equation}
Here, \(\omega_m\) is the angular frequency of the effective phonon or mechanical mode, \(a^{\dagger}\) is the phonon creation operator, and \(a\) is the phonon annihilation operator. The number operator \(a^{\dagger}a\) gives the number of phonons in the mode. If the phonon mode is in the Fock state \(\left|n\right\rangle\), then
\begin{equation}
a^{\dagger}a\left|n\right\rangle=n\left|n\right\rangle .
\end{equation}
The corresponding phonon energy is
\begin{equation}
E_n=\hbar \omega_m n .
\end{equation}
The zero-point energy term \((1/2)\hbar \omega_m\) is omitted because it adds only a constant energy shift and does not affect the system dynamics. Including \(H_{ph}\) allows the model to describe the effect of lattice vibrations on the NV-center spin system.

\subsection{Effective Electron--Phonon Coupling}

The electron--phonon interaction is modeled using an effective longitudinal coupling term motivated by spin-strain and spin-lattice coupling in NV centers \cite{macquarrie2013,cambria2023,thiering2025}:
\begin{equation}
H_{e-ph}=g_{ph}S_z\left(a+a^{\dagger}\right).
\end{equation}
Here, \(g_{ph}\) is the effective electron--phonon coupling strength. The term \(a+a^{\dagger}\) represents the displacement quadrature of the phonon mode, while \(S_z\) represents the electron-spin projection along the NV axis. This interaction describes how lattice vibrations modulate the energy splitting of the NV electron spin.

Physically, the diamond lattice surrounding the NV center is not perfectly static. Vibrational motion of the lattice changes the local strain environment of the defect. Since the NV spin levels are sensitive to strain, phonon-induced lattice displacement can shift the electron-spin transition frequency. These fluctuations produce phase noise and relaxation channels for the electron spin. Because the electron spin is actively involved in the write and read stages of the memory protocol, phonon-induced perturbations can reduce the fidelity of electron-to-nuclear spin state transfer.

In the present work, \(g_{ph}\) is treated as an effective coupling parameter. The purpose of this term is not to reproduce the complete microscopic spin-strain Hamiltonian of the NV center, but to capture the influence of phonon-induced perturbations on quantum memory performance. A more complete treatment can replace this simplified interaction with the full symmetry-allowed spin-strain Hamiltonian of the NV center \cite{doherty2013,macquarrie2013}. However, the effective form used here is sufficient to study how increasing electron--phonon coupling strength influences memory fidelity.

\subsection{External Control Hamiltonian}

The term \(H_{drive}\) represents the external control fields applied to the NV electron and nuclear spins. In experiments, microwave fields are commonly used to drive electron-spin transitions, while radio-frequency fields are used to manipulate nuclear-spin transitions. A general time-dependent drive Hamiltonian may be written as \cite{jelezko2006,childress2013}
\begin{equation}
H_{drive}(t)=\hbar \Omega_e \cos(\omega_e t)S_x+\hbar \Omega_n \cos(\omega_n t)I_x,
\end{equation}
where \(\Omega_e\) and \(\Omega_n\) are the electron and nuclear drive strengths, and \(\omega_e\) and \(\omega_n\) are the corresponding drive frequencies.

For the purpose of modeling the memory transfer process, the detailed pulse sequence can be represented by an effective rotating-frame SWAP-like Hamiltonian, consistent with hyperfine-mediated electron-to-nuclear transfer models \cite{fuchs2011,shim2013}:
\begin{equation}
\begin{split}
H_{swap}=\hbar\Omega\big(&\left|0\right\rangle_e\left|1\right\rangle_n\,{}_e\!\left\langle 1\right|{}_n\!\left\langle 0\right| \\
&+\left|1\right\rangle_e\left|0\right\rangle_n\,{}_e\!\left\langle 0\right|{}_n\!\left\langle 1\right|\big),
\end{split}
\end{equation}
where \(\Omega\) is the effective electron--nuclear transfer rate. This Hamiltonian couples the states \(\left|1\right\rangle_e\left|0\right\rangle_n\) and \(\left|0\right\rangle_e\left|1\right\rangle_n\), thereby enabling transfer of quantum information from the electron spin to the nuclear spin. In this effective description, \(H_{drive}\) activates the write and read operations of the quantum memory.

\subsection{Total Effective Hamiltonian for Memory Simulation}

Combining the above terms, the total phonon-aware Hamiltonian is
\begin{equation}
\begin{split}
H={}&D S_z^2+\gamma_e B_z S_z-\gamma_n B_z I_z+A_{\parallel}S_z I_z+A_{\perp}\left(S_x I_x+S_y I_y\right)\\
&+\hbar\omega_m a^{\dagger}a+g_{ph}S_z\left(a+a^{\dagger}\right)+H_{drive}.
\end{split}
\end{equation}
This Hamiltonian captures the key coherent interactions relevant to the proposed quantum memory model. The electron and nuclear spin Hamiltonians define the individual qubit energy structures. The hyperfine interaction couples the electron and nuclear spins and enables state transfer. The phonon Hamiltonian represents lattice vibrations, while the electron--phonon interaction introduces phonon-induced perturbations to the electron spin. The drive Hamiltonian implements the external control needed for write and read operations.

For numerical simulation, it is often convenient to work in an effective rotating frame. In this frame, the large zero-field splitting term can be removed from the fast dynamics, while the secular part of the hyperfine coupling, the phonon interaction, and the control Hamiltonian are retained; the corresponding approximations are stated in Section~\ref{sec:simulation}. This avoids unnecessary numerical oscillations at the gigahertz scale and allows the memory dynamics to be studied on the relevant transfer and storage time scales.

The proposed Hamiltonian therefore provides a physically motivated model for analyzing electron-to-nuclear spin quantum memory in the presence of phonon-induced decoherence.

\section{Theoretical Justification of the Proposed Memory Model}

After defining the system Hamiltonian, the next step is to show theoretically that the proposed model can perform electron-to-nuclear spin quantum memory and to explain how phonons reduce the memory fidelity. This section provides the analytical basis for the proposed memory protocol.

\subsection{Effective Two-Qubit Memory Subspace}

The NV electron spin has three ground-state spin sublevels. However, for memory operation, only two of these levels are selected to form an effective electron-spin qubit:
\begin{equation}
\left|0\right\rangle_e \equiv \left|m_s=0\right\rangle,\qquad
\left|1\right\rangle_e \equiv \left|m_s=-1\right\rangle .
\end{equation}
The nuclear spin is used as the memory qubit and is represented by
\begin{equation}
\left|0\right\rangle_n,\qquad \left|1\right\rangle_n .
\end{equation}
An arbitrary electron-spin qubit state can be written in the standard pure-state form \cite{nielsen2010}:
\begin{equation}
\left|\psi_e\right\rangle=\alpha\left|0\right\rangle_e+\beta\left|1\right\rangle_e,
\end{equation}
where \(\alpha\) and \(\beta\) are complex probability amplitudes satisfying
\begin{equation}
\left|\alpha\right|^2+\left|\beta\right|^2=1 .
\end{equation}
The nuclear spin is initially prepared in the state \(\left|0\right\rangle_n\). Therefore, the initial combined electron--nuclear state is
\begin{equation}
\left|\Psi(0)\right\rangle=\left(\alpha\left|0\right\rangle_e+\beta\left|1\right\rangle_e\right)\left|0\right\rangle_n .
\end{equation}
The objective of the write operation is to transfer the quantum information from the electron spin to the nuclear spin, such that
\begin{equation}
\left(\alpha\left|0\right\rangle_e+\beta\left|1\right\rangle_e\right)\left|0\right\rangle_n
\rightarrow
\left|0\right\rangle_e\left(\alpha\left|0\right\rangle_n+\beta\left|1\right\rangle_n\right).
\end{equation}
This transformation represents electron-to-nuclear spin quantum memory \cite{fuchs2011,shim2013}. The electron spin is returned to a known state \(\left|0\right\rangle_e\), while the unknown quantum amplitudes \(\alpha\) and \(\beta\) are stored in the nuclear spin.

\subsection{Ideal Electron-to-Nuclear State Transfer}

The effective transfer operation is described using the SWAP-like Hamiltonian introduced above, which is an idealized representation of controlled electron--nuclear state exchange \cite{fuchs2011,shim2013}:
\begin{equation}
\begin{split}
H_{swap}=\hbar\Omega\big(&\left|0\right\rangle_e\left|1\right\rangle_n\,{}_e\!\left\langle 1\right|{}_n\!\left\langle 0\right| \\
&+\left|1\right\rangle_e\left|0\right\rangle_n\,{}_e\!\left\langle 0\right|{}_n\!\left\langle 1\right|\big).
\end{split}
\end{equation}
This Hamiltonian acts only in the two-dimensional subspace spanned by
\begin{equation}
\left|1\right\rangle_e\left|0\right\rangle_n
\end{equation}
and
\begin{equation}
\left|0\right\rangle_e\left|1\right\rangle_n .
\end{equation}
It couples these two states while leaving \(\left|0\right\rangle_e\left|0\right\rangle_n\) unchanged. Under this Hamiltonian, the state \(\left|1\right\rangle_e\left|0\right\rangle_n\) evolves as
\begin{equation}
\left|1\right\rangle_e\left|0\right\rangle_n
\rightarrow
\cos(\Omega t)\left|1\right\rangle_e\left|0\right\rangle_n-i\sin(\Omega t)\left|0\right\rangle_e\left|1\right\rangle_n .
\end{equation}
For an interaction time
\begin{equation}
t_{swap}=\frac{\pi}{2\Omega},
\end{equation}
the evolution becomes
\begin{equation}
\left|1\right\rangle_e\left|0\right\rangle_n
\rightarrow
-i\left|0\right\rangle_e\left|1\right\rangle_n .
\end{equation}
Since the state \(\left|0\right\rangle_e\left|0\right\rangle_n\) is not affected by this Hamiltonian, the full initial state evolves as
\begin{equation}
\left(\alpha\left|0\right\rangle_e+\beta\left|1\right\rangle_e\right)\left|0\right\rangle_n
\rightarrow
\left|0\right\rangle_e\left(\alpha\left|0\right\rangle_n-i\beta\left|1\right\rangle_n\right).
\end{equation}
This result shows that the quantum information originally encoded in the electron spin has been transferred to the nuclear spin. The factor \(-i\) is a deterministic phase acquired during the coherent exchange process. Since this phase is known, it can be corrected using a single-qubit phase operation on the nuclear spin. Therefore, in the absence of decoherence, the proposed Hamiltonian realizes an ideal electron-to-nuclear memory write operation.

\subsection{Storage and Read Operation}

After the write operation, the electron spin is in the known state \(\left|0\right\rangle_e\), while the nuclear spin stores the quantum state
\begin{equation}
\left|\psi_n\right\rangle=\alpha\left|0\right\rangle_n-i\beta\left|1\right\rangle_n .
\end{equation}
During the storage interval \(T_s\), the nuclear spin preserves the quantum information. Since nuclear spins are generally less sensitive to environmental noise than electron spins, they are suitable for long-lived quantum memory.

The read operation is performed by applying the reverse transfer process, which maps the nuclear-spin state back to the electron spin. Ideally, after the complete write--store--read sequence and phase correction, the final state becomes
\begin{equation}
\left|\Psi_{\mathrm{target}}\right\rangle=
\left(\alpha\left|0\right\rangle_e+\beta\left|1\right\rangle_e\right)\left|0\right\rangle_n .
\end{equation}
Thus, the complete memory protocol consists of three stages:
\begin{equation}
\text{Write}\rightarrow\text{Store}\rightarrow\text{Read}.
\end{equation}
The write stage transfers the electron-spin state to the nuclear spin. The storage stage preserves the state in the nuclear-spin memory. The read stage transfers the stored state back to the electron spin for measurement.

\subsection{Effect of Phonon-Induced Dephasing}

In an ideal closed system, the above transfer process is unitary and reversible. However, in a real diamond lattice, the NV center interacts with phonons. The effective electron--phonon coupling is written as
\begin{equation}
H_{e-ph}=g_{ph}S_z\left(a+a^{\dagger}\right).
\end{equation}
This interaction causes fluctuations in the electron-spin energy splitting because the phonon displacement modifies the local strain environment of the NV center \cite{macquarrie2013,cambria2023}. These energy fluctuations produce phase noise in the electron-spin qubit. Since the electron spin participates directly in the write and read stages, phonon-induced electron-spin dephasing reduces the quality of the state-transfer operation.

The effect of phonon-induced dephasing can be represented by the decay of the off-diagonal coherence term in the electron-spin density matrix:
\begin{equation}
\rho_{01}(t)=\rho_{01}(0)e^{-\chi_{ph}(t)} .
\end{equation}
Here, \(\rho_{01}(t)\) represents the electron-spin coherence between \(\left|0\right\rangle_e\) and \(\left|1\right\rangle_e\), and \(\chi_{ph}(t)\) is the phonon-induced dephasing function. The function \(\chi_{ph}(t)\) depends on the electron--phonon coupling strength, phonon spectral density, phonon frequency, and thermal phonon occupation.

For a phonon bath, a general dephasing function may be written in the form used in standard open-system treatments of bosonic environments \cite{breuer2002}:
\begin{equation}
\chi_{ph}(t)=\int_{0}^{\infty} J(\omega)\coth\left(\frac{\hbar\omega}{2k_B T}\right)\frac{1-\cos(\omega t)}{\omega^2}\,d\omega,
\end{equation}
where \(J(\omega)\) is the phonon spectral density, \(T\) is the temperature, and \(k_B\) is Boltzmann's constant. Since \(J(\omega)\geq 0\), \(\coth(\hbar\omega/2k_B T)\geq 1\), and \(1-\cos(\omega t)\geq 0\), the dephasing function satisfies
\begin{equation}
\chi_{ph}(t)\geq 0 .
\end{equation}
Therefore,
\begin{equation}
e^{-\chi_{ph}(t)}\leq 1 .
\end{equation}
This means that phonon coupling can only preserve or reduce electron-spin coherence in this model. It cannot improve the ideal coherence. As the spin-phonon coupling strength or the temperature increases, \(\chi_{ph}(t)\) increases, causing stronger coherence loss.

\subsection{Temperature-Dependent Phonon Occupation}

The average number of thermal phonons in a mode of angular frequency \(\omega_m\) is given by the Bose--Einstein distribution:
\begin{equation}
\bar{n}(T)=\frac{1}{\exp\left(\frac{\hbar\omega_m}{k_B T}\right)-1} .
\end{equation}
At low temperature, \(k_B T\ll \hbar\omega_m\), the thermal phonon occupation becomes very small:
\begin{equation}
\bar{n}(T)\approx 0 .
\end{equation}
At higher temperature, \(k_B T\gg \hbar\omega_m\), the phonon occupation increases approximately as
\begin{equation}
\bar{n}(T)\approx\frac{k_B T}{\hbar\omega_m}.
\end{equation}
Therefore, increasing the temperature increases the number of thermally occupied phonons. This increases phonon-induced fluctuations in the electron-spin transition frequency and enhances dephasing. As a result, the memory fidelity is expected to decrease with increasing temperature.

\subsection{Memory Fidelity Expression}

The quality of the memory operation is quantified using the memory fidelity \cite{nielsen2010}
\begin{equation}
F_{\mathrm{mem}}=\left\langle\Psi_{\mathrm{target}}\right|\rho_{\mathrm{out}}\left|\Psi_{\mathrm{target}}\right\rangle,
\end{equation}
where \(\rho_{\mathrm{out}}\) is the final density matrix obtained after the write--store--read protocol, and \(\left|\Psi_{\mathrm{target}}\right\rangle\) is the ideal final state.

In a simplified dephasing-dominated model, the fidelity can be approximated as
\begin{equation}
F_{\mathrm{mem}}\approx\frac{1}{2}\left[1+\exp(-\Lambda)\right],
\end{equation}
where \(\Lambda\) represents the total decoherence accumulated during the memory operation. A useful expression for \(\Lambda\) is
\begin{equation}
\Lambda=2\Gamma_{\phi e}t_{swap}+\Gamma_{\phi n}T_s+\chi_{ph}(2t_{swap}).
\end{equation}
Here, \(\Gamma_{\phi e}\) is the electron-spin dephasing rate, \(\Gamma_{\phi n}\) is the nuclear-spin dephasing rate, \(t_{swap}\) is the duration of the electron--nuclear transfer operation, and \(T_s\) is the nuclear-spin storage time. The factor \(2t_{swap}\) appears because the electron spin is involved in both the write and read stages. The term \(\chi_{ph}(2t_{swap})\) represents the phonon-induced dephasing accumulated during the electron-mediated portions of the protocol.

In the ideal case,
\begin{equation}
\Gamma_{\phi e}=\Gamma_{\phi n}=\chi_{ph}(t)=0 .
\end{equation}
Then,
\begin{equation}
\Lambda=0
\end{equation}
and
\begin{equation}
F_{\mathrm{mem}}=1 .
\end{equation}
This corresponds to perfect quantum memory.

However, when phonon-induced dephasing is present,
\begin{equation}
\chi_{ph}(t)>0,
\end{equation}
which gives
\begin{equation}
\Lambda>0
\end{equation}
and therefore
\begin{equation}
F_{\mathrm{mem}}<1 .
\end{equation}
This result provides the theoretical basis for the proposed phonon-aware memory model. It shows that stronger electron--phonon coupling, higher thermal phonon occupation, and longer storage time reduce the electron-to-nuclear spin memory fidelity.

\subsection{Theoretical Claim of the Proposed Model}

The theoretical analysis establishes two important points. First, in the absence of decoherence, the effective electron--nuclear SWAP-like Hamiltonian transfers an arbitrary electron-spin qubit state to the nuclear-spin memory qubit, apart from a known and correctable phase. Second, when electron--phonon coupling is included, phonon-induced dephasing suppresses electron-spin coherence during the write and read stages, thereby reducing the total memory fidelity.

Therefore, the proposed model theoretically supports the main claim of this work:
\begin{equation}
g_{ph}\uparrow \Rightarrow \chi_{ph}(t)\uparrow \Rightarrow F_{\mathrm{mem}}\downarrow .
\end{equation}
Similarly,
\begin{equation}
T\uparrow \Rightarrow \bar{n}(T)\uparrow \Rightarrow \chi_{ph}(t)\uparrow \Rightarrow F_{\mathrm{mem}}\downarrow .
\end{equation}
Also,
\begin{equation}
T_s\uparrow \Rightarrow \Gamma_{\phi n}T_s\uparrow \Rightarrow F_{\mathrm{mem}}\downarrow .
\end{equation}
Thus, the theoretical model predicts that memory performance is limited by electron-spin decoherence during write and read operations, nuclear-spin decoherence during storage, and phonon-induced dephasing caused by lattice vibrations. These theoretical predictions are validated numerically in the following simulation section using an open quantum system model based on the Lindblad master equation.

\section{QuTiP Simulation Methodology and Results}
\label{sec:simulation}

The theoretical model was implemented as a conceptual open-system simulation in QuTiP \cite{johansson2012,johansson2013}. The simulation uses the tensor-product Hilbert space
\begin{equation}
\mathcal{H}=\mathcal{H}_e\otimes \mathcal{H}_n\otimes \mathcal{H}_{ph},
\end{equation}
where the NV electron spin is represented as a three-level spin-1 system, the nuclear-spin memory is represented as a two-level spin-\(1/2\) system, and the phonon mode is represented by a truncated harmonic-oscillator basis with eight Fock states. The electron qubit is encoded using \(\left|m_s=0\right\rangle\) and \(\left|m_s=-1\right\rangle\), while the nuclear qubit is initialized in \(\left|0\right\rangle_n\). The initial electron state is chosen as an equal superposition with \(\alpha=\beta=1/\sqrt{2}\), so that the simulation directly tests the preservation of quantum coherence during the write--store--read process.

The numerical implementation uses microseconds as the time unit and megahertz as the frequency unit. All Hamiltonian frequencies are converted to angular frequencies using \(\omega=2\pi f\). During the write and read stages, the effective rotating-frame Hamiltonian is taken as
\begin{equation}
H_{wr}=H_{ph}+H_{e-ph}+H_{swap},
\end{equation}
whereas during the storage stage the SWAP drive is turned off and the evolution is governed by
\begin{equation}
H_{s}=H_{ph}+H_{e-ph}.
\end{equation}
Two standard approximations lead to this form. First, the transverse hyperfine term \(A_{\perp}(S_xI_x+S_yI_y)\) connects electron levels separated by the zero-field splitting \(D\approx 2.87\,\mathrm{GHz}\), which exceeds \(A_{\perp}\) by more than four orders of magnitude. In the electron rotating frame this term is therefore non-secular and is neglected; its residual effect scales as \((A_{\perp}/D)^2\sim10^{-10}\) for the present parameters. Second, the secular term \(A_{\parallel}S_zI_z\) commutes with \(H_{ph}\), \(H_{e-ph}\), and the dephasing operators, and it only changes the phases of the relaxation operators, which leaves the Lindblad dissipator unchanged. It can therefore be removed exactly by moving to its interaction picture, in which the microwave/radio-frequency SWAP drive, calibrated to the hyperfine-shifted transition as in experiments, becomes resonant. Retaining both hyperfine terms in the rotating frame without \(D\) and without drive compensation would instead produce spurious coherent electron--nuclear oscillations during storage and a detuned, incomplete SWAP operation.
For the selected effective transfer coupling \(J_{swap}=0.20\,\mathrm{MHz}\), the corresponding SWAP time used in the notebook is
\begin{equation}
t_{swap}=\frac{1}{4J_{swap}}=1.25\,\mu\mathrm{s}.
\end{equation}
The open-system evolution is solved using the Lindblad master equation \cite{lindblad1976,gorini1976,breuer2002},
\begin{equation}
\frac{d\rho}{dt}=-i[H,\rho]+\sum_k\left(c_k\rho c_k^{\dagger}-\frac{1}{2}\left\{c_k^{\dagger}c_k,\rho\right\}\right),
\end{equation}
where the collapse operators represent electron relaxation, electron pure dephasing, nuclear relaxation, nuclear dephasing, and phonon damping. The phonon bath occupation is evaluated using the Bose--Einstein thermal occupation model \cite{breuer2002}. Since the present simulation is conceptual rather than fitted to a specific experiment, the absolute fidelity values should be interpreted as model-level indicators of the influence of the chosen parameters.

\begin{figure}[htbp]
  \centering
  \includegraphics[width=0.70\textwidth]{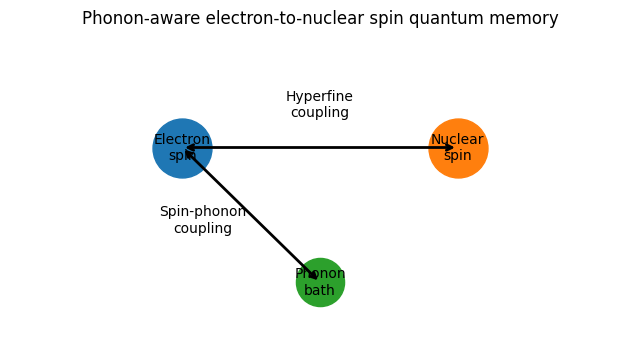}
  \caption{Conceptual simulation model showing the electron spin, nuclear-spin memory, hyperfine coupling, and phonon bath.}
\end{figure}

\begin{figure}[htbp]
  \centering
  \includegraphics[width=0.70\textwidth]{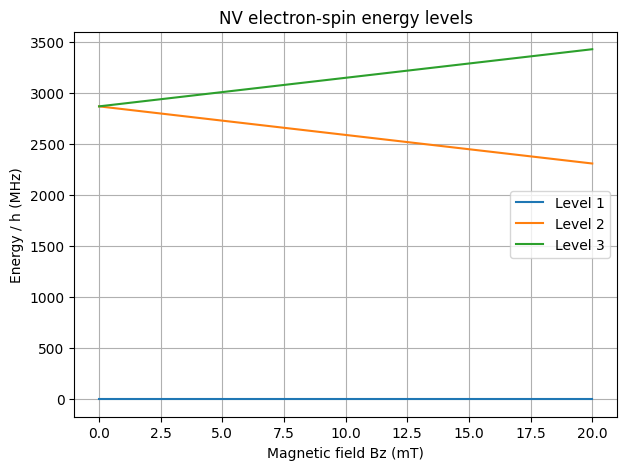}
  \caption{Simulated NV electron-spin energy levels as a function of axial magnetic field. The zero-field splitting and Zeeman splitting define the electron-spin transition structure used for qubit selection.}
\end{figure}

\begin{figure}[htbp]
  \centering
  \includegraphics[width=0.85\textwidth]{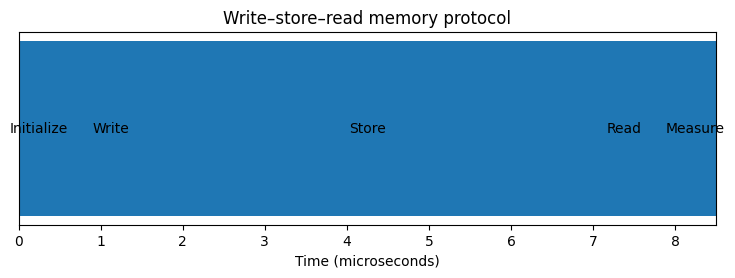}
  \caption{Write--store--read memory protocol used in the simulation. The write and read windows correspond to the effective SWAP duration of \(1.25\,\mu\mathrm{s}\).}
\end{figure}

\subsection{Single-Run Output}

A representative single simulation run was performed with a storage time of \(10\,\mu\mathrm{s}\) and effective spin-phonon coupling \(g_{ph}=0.02\,\mathrm{MHz}\). The write fidelity obtained after the electron-to-nuclear transfer was \(0.9994\). After the full write--store--read sequence, the final memory fidelity was \(0.9982\), the transfer efficiency was \(0.9994\), and the final reduced spin purity was \(0.9964\). The corresponding thermal phonon occupation used in the run was \(\bar{n}=0.6206\). The residual infidelity arises mainly from electron--phonon entanglement created while the electron occupies \(\left|m_s=-1\right\rangle\) during the write and read stages, which also lowers the purity of the reduced spin state.

\subsection{Storage-Time and Spin-Phonon-Coupling Analysis}

\begin{figure}[htbp]
  \centering
  \includegraphics[width=0.92\textwidth]{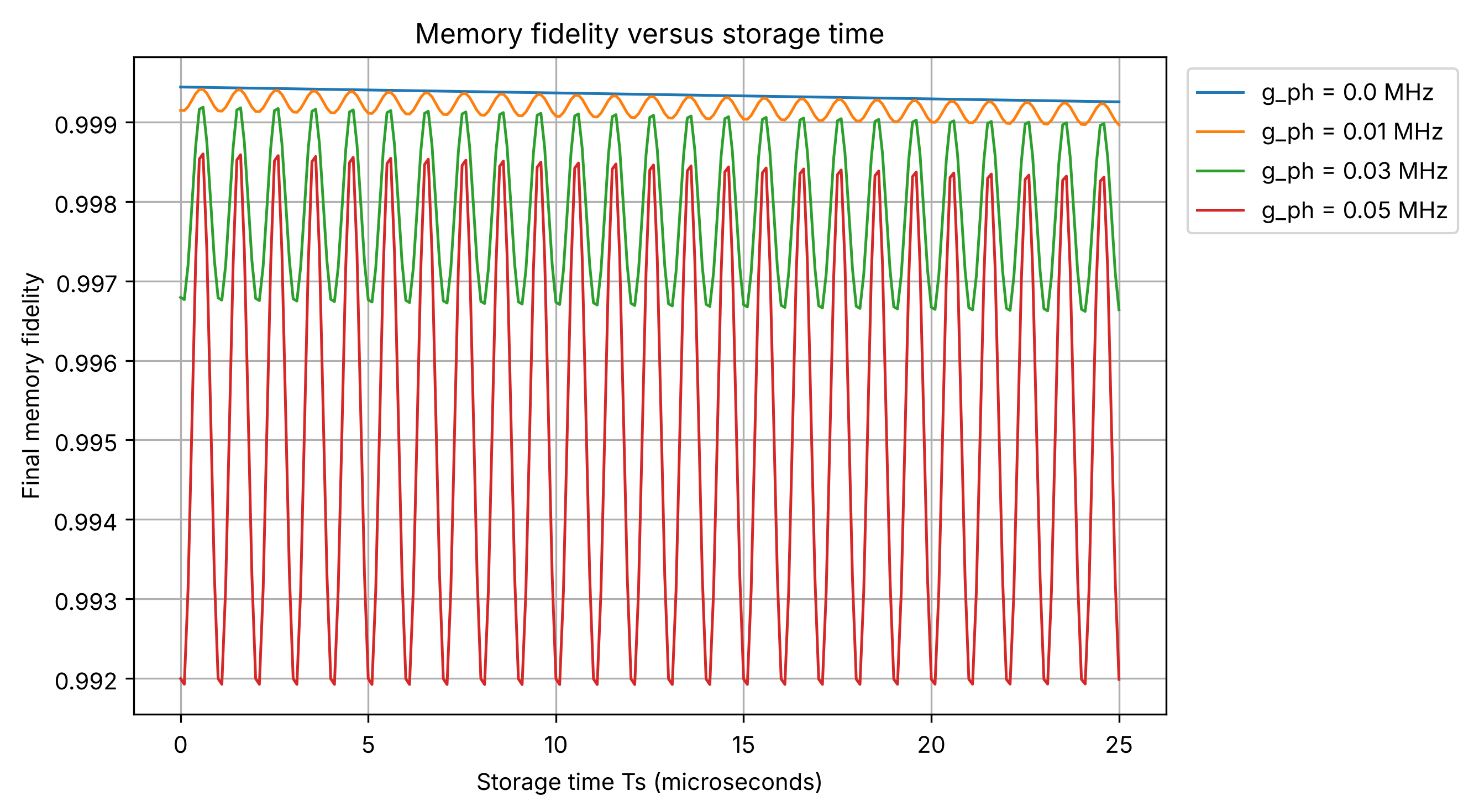}
  \caption{Final memory fidelity versus nuclear-spin storage time for different spin-phonon coupling strengths. The oscillation period equals the phonon-mode period \(1/f_m=1\,\mu\mathrm{s}\).}
  \label{fig:fidelity_storage}
\end{figure}

Figure~\ref{fig:fidelity_storage} shows the final memory fidelity as a function of storage time for different values of the effective spin-phonon coupling strength \(g_{ph}\), sampled every \(0.1\,\mu\mathrm{s}\). Without spin-phonon coupling, the fidelity decays slowly and almost linearly, by \(7.5\times10^{-6}\) per microsecond. This matches the rate \(\Gamma_{\phi n}/2+\Gamma_{1n}/4\) expected for an equal superposition stored under the chosen nuclear relaxation and dephasing rates.

With \(g_{ph}>0\), the fidelity oscillates with the phonon-mode period \(1/f_m=1\,\mu\mathrm{s}\). During the write step, the \(\left|m_s=-1\right\rangle\) component displaces the phonon mode and becomes entangled with it. During storage, the electron is in \(\left|m_s=0\right\rangle\), which does not couple to the phonon mode, so the mode evolves freely. Whether the read step undoes or reinforces the displacement then depends on the phonon phase at the start of the read. The lower edge of the oscillation decreases with coupling strength, from \(0.9994\) at \(g_{ph}=0\) to \(0.9991\), \(0.9967\), and \(0.9918\) for \(g_{ph}=0.01\), \(0.03\), and \(0.05\,\mathrm{MHz}\), while the upper edge remains above \(0.9986\). These revivals are a feature of the single-mode phonon description. A continuum of phonon modes would wash them out, so the lower envelope is the more conservative estimate of memory performance.

\subsection{Temperature-Dependent Simulation}

\begin{figure}[htbp]
  \centering
  \includegraphics[width=0.75\textwidth]{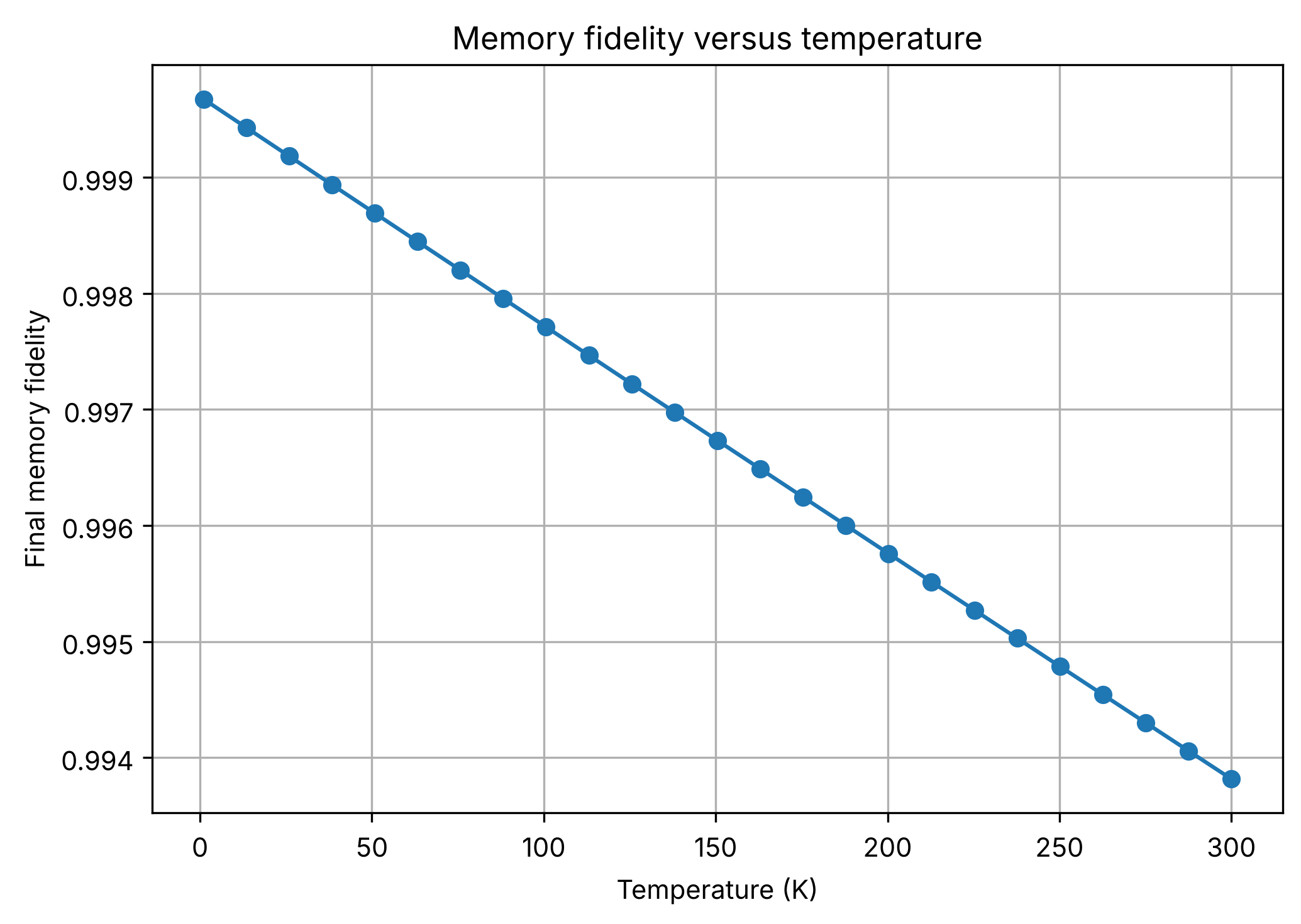}
  \caption{Final memory fidelity versus temperature using the effective temperature-dependent Lindblad-rate model in the notebook.}
  \label{fig:fidelity_temperature}
\end{figure}

The notebook also evaluates a temperature-dependent effective-rate model. In this model, temperature modifies the electron relaxation and dephasing rates through a phenomenological phonon-occupation factor, rather than by explicitly increasing the phonon Hilbert-space size. Figure~\ref{fig:fidelity_temperature} shows the corresponding final memory fidelity for temperatures from \(1\,\mathrm{K}\) to \(300\,\mathrm{K}\). The fidelity decreases monotonically, from \(0.9997\) at \(1\,\mathrm{K}\) to \(0.9982\) at \(77\,\mathrm{K}\) and \(0.9938\) at \(300\,\mathrm{K}\). This is consistent with the theoretical expectation that higher thermal phonon occupation increases electron relaxation and dephasing during the write and read stages. Because the electron participates only for \(2t_{swap}=2.5\,\mu\mathrm{s}\), the overall reduction remains small for the chosen rates. The absolute values depend on the phenomenological rate parameters, which should be fitted to measured NV spin-lattice relaxation data \cite{cambria2023} for quantitative predictions.

\subsection{Two-Parameter Fidelity Map}

\begin{figure}[htbp]
  \centering
  \includegraphics[width=0.75\textwidth]{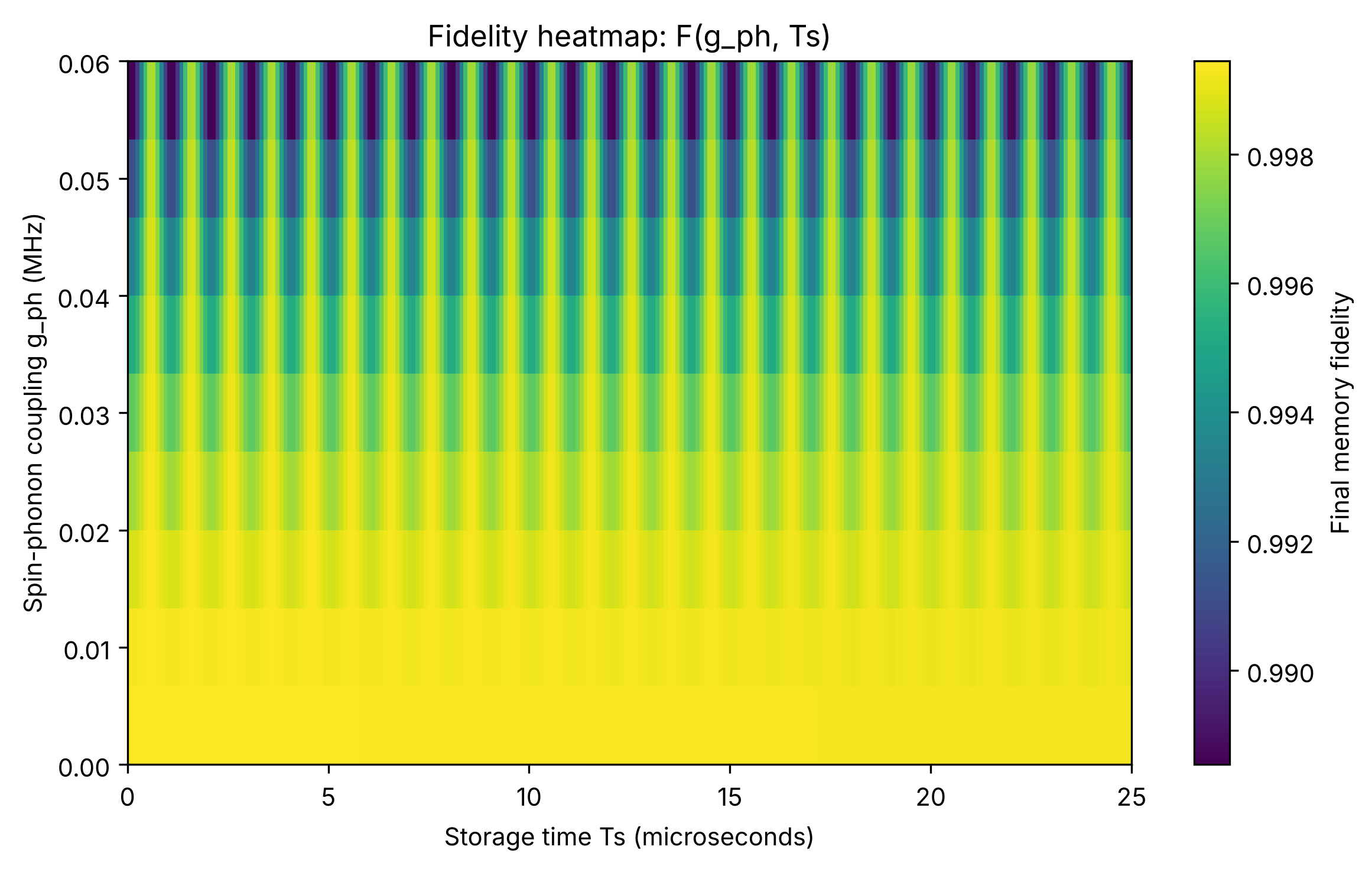}
  \caption{Heatmap of final memory fidelity as a function of storage time and spin-phonon coupling strength.}
  \label{fig:fidelity_heatmap}
\end{figure}

Figure~\ref{fig:fidelity_heatmap} summarizes the combined dependence of memory fidelity on storage time \(T_s\) and spin-phonon coupling strength \(g_{ph}\). The fidelity decreases along the coupling axis, from about \(0.999\) for \(g_{ph}\leq0.01\,\mathrm{MHz}\) to a minimum of \(0.9884\) at \(g_{ph}=0.06\,\mathrm{MHz}\), and the vertical stripes reflect the phonon-period oscillation discussed above. Along the storage-time axis, losses within \(25\,\mu\mathrm{s}\) are small compared with the phonon-induced reduction, because nuclear-spin decoherence is slow for the chosen rates. In this model, spin-phonon coupling during the electron-mediated write and read stages is therefore the dominant source of memory error.

\subsection{Summary of Numerical Results}

Table~\ref{tab:simulation_results} summarizes the main numerical outputs generated by the notebook. The ideal-memory case removes electron relaxation, electron dephasing, phonon damping, and spin-phonon coupling, and reaches \(F_f=0.99997\), which confirms that the transfer protocol and phase correction are implemented correctly. The write fidelity and transfer efficiency remain above \(0.997\) in all cases. Strong spin-phonon coupling (\(g_{ph}=0.05\,\mathrm{MHz}\)) lowers the final fidelity to \(0.9920\) and the spin purity to \(0.9844\). By contrast, extending the storage time from \(10\,\mu\mathrm{s}\) to \(30\,\mu\mathrm{s}\) at weak coupling reduces the final fidelity only from \(0.9991\) to \(0.9989\).

\begin{table}[htbp]
\caption{Numerical results obtained from the QuTiP notebook.}
\label{tab:simulation_results}
\centering
\small
\begin{tabular}{p{0.24\textwidth}cccccc}
\toprule
Case & \(g_{ph}\) & \(T_s\) & \(F_w\) & \(F_f\) & \(\eta\) & Purity \\
 & (MHz) & (\(\mu\mathrm{s}\)) & & & & \\
\midrule
Ideal memory & 0.00 & 2.0 & 1.0000 & 1.0000 & 1.0000 & 0.9999 \\
Weak phonon coupling & 0.01 & 10.0 & 0.9996 & 0.9991 & 0.9995 & 0.9982 \\
Strong phonon coupling & 0.05 & 10.0 & 0.9977 & 0.9920 & 0.9988 & 0.9844 \\
Long storage & 0.01 & 30.0 & 0.9996 & 0.9989 & 0.9995 & 0.9979 \\
\bottomrule
\end{tabular}
\end{table}

Overall, the simulation supports the usefulness of the proposed phonon-aware open-system framework implemented using QuTiP \cite{johansson2012,johansson2013}. It shows how the electron-spin write/read operation, nuclear-spin storage interval, spin-phonon coupling, and Lindblad decoherence channels can be evaluated within one reproducible QuTiP workflow. At the same time, the model remains conceptual: the absolute fidelities depend on the selected phenomenological rates and on the single-mode phonon description. The next step is therefore to calibrate the relaxation and dephasing rates, the spin-phonon coupling, and the phonon spectral density using experimental NV-center parameters, and to replace the single mode with a continuum phonon bath, before claiming quantitative agreement with measured quantum-memory performance.

\section{Conclusion}

This work presented a theoretical and simulation-based model for phonon-aware electron-to-nuclear spin quantum memory in NV centers in diamond. The model treats the NV electron spin as a controllable interface qubit and a nearby nuclear spin as a long-lived memory qubit, with hyperfine-mediated, drive-assisted state transfer between them, and adds an effective phonon mode coupled to the electron spin. Simulations in the secular, hyperfine-resonant rotating frame show that the write--store--read protocol is essentially lossless without decoherence. For the chosen parameters, spin-phonon coupling during the electron-mediated write and read stages is the dominant error source: increasing \(g_{ph}\) to \(0.06\,\mathrm{MHz}\) lowers the worst-case fidelity to \(0.9884\), with revivals at the phonon-mode period, and higher temperature lowers the fidelity monotonically through faster electron relaxation and dephasing. Nuclear-spin storage itself contributes only a slowly accumulating error. These results support treating phonon effects, and in particular the duration of the electron-mediated steps, as an important design factor for NV-center quantum memories. Future work on this model can include experimentally fitted spin--phonon parameters, a continuum phonon bath, full spin--strain Hamiltonians, multiple nuclear-spin registers, and optimized control pulses to improve memory robustness.

\section*{Code Availability}
The QuTiP simulation code, the notebook, and the scripts used to generate all figures and Table~\ref{tab:simulation_results} are openly available at \url{https://github.com/manjithbc/nv-quantum-memory} under the MIT License.

\end{document}